\documentstyle [twoside,12pt]{article}
\begin{document}
\setlength{\baselineskip}{.7cm}
\sloppy
\begin{center}
{\bf EXACT SOLUTIONS FOR WAVE PROPAGATION IN BIREFRINGENT OPTICAL FIBERS}
\end{center}
\begin{center}
{\bf E. Alfinito, M. Leo, R.A. Leo, G. Soliani and L. Solombrino }
\end{center}
\begin{center}
{\it Dipartimento di Fisica dell'Universit\`a, 73100 Lecce, Italy\\
 and Istituto Nazionale di Fisica Nucleare, Sezione di Lecce. }
\end{center}
\medskip
\medskip
\medskip
\medskip
\begin{abstract}
We carry out a group-theoretical study of the pair of nonlinear
Schr\"{o}dinger equations describing the propagation of waves in nonlinear
birefringent optical fibers. We exploit the symmetry algebra
associated with these equations to provide examples of
specific exact solutions. Among them, we obtain the soliton profile, which
is related to the coordinate translations and the
constant change of phase.
\end{abstract}
\bigskip
\section{Introduction}
The propagation of optical pulses in nonlinear birefringent fibers is
described by the pair of nonlinear Schr\"{o}dinger equations
${\displaystyle{\cite{LO}}}$
$$ \Delta_{1}= i\:u_{x}+u_{tt}+kv+(\alpha |u|^{2}+\beta|v|^{2})u=0, \eqno(1.
1a)$$
$$ \Delta_{2}= i\:v_{x}+v_{tt}+ku+(\alpha |v|^{2}+\beta|u|^{2})v=0, \eqno(1.
1b)$$
where $u=u(x,t)$ and $v=v(x,t)$ are the circularly polarized components of
the optical field, $x$ and $t$ denote the (normalized) longitudinal coordinate
of the
fiber and the time variable, respectively, $k$ is the birefringence
parameter and the coefficients $\alpha$ and $\beta$ are responsible for the
nonlinear properties of the fiber ${\displaystyle{\cite{TR,ST,CA}}}$.
Performing the change of variables
$x \longrightarrow \alpha x, \quad t \longrightarrow \mu t, \quad {\rm
with}\;\; \mu^{2} = \pm\frac{1}{2}\alpha,$   Eqs.(1.1) take the form
$$ i\:u_{x}\pm \frac{1}{2}u_{tt}+k^{\prime}v+( |u|^{2}+\sigma|v|^{2})u=0,
\eqno(1.
2a)$$
$$ i\:v_{x}\pm \frac{1}{2}v_{tt}+k^{\prime}u+ (|v|^{2}+\sigma|u|^{2})v=0,
\eqno(1.
2b)$$
where $k^{\prime}=k/\alpha,$ + (-) holds in the anomalous (normal) dispersion
regime and $\sigma=
\frac{\beta}{\alpha}= \frac{1+B}{1-B},$ $B$ being the third-order
susceptibility coefficient ${\displaystyle{\cite{TR,ST,CA}}}$.
For $\alpha=\beta\;\;\; (\sigma=1)$ and $k=0$, the system (1.1) has an infinite
set
of constants of motion and may be solved by the inverse scattering method
${\displaystyle{\cite{ZA}}}$.

In general, i.e. for $\alpha \ne \beta$ and $k \ne 0$, the system (1.1) is
not integrable by inverse scattering. In this case, Eqs.(1.1) possess
three constants of motion only ${\displaystyle{\cite{LO}}}$. On
the other hand, it is well-known that a powerful tool for handling both
integrable and nonintegrable differential equations is represented by the
so-called symmetry approach ${\displaystyle{\cite{OL}}}$. This method,
which is based on the Lie group theory, consists essentially in looking for
symmetry transformations that reduce the equations under consideration to
certain ordinary differential equations, each of them comes from an
invariant quantity associated with a given symmetry allowed by the system.
Following this idea, in this work we apply the symmetry approach to
Eqs.(1.1). We display examples of new exact solutions in both the cases
$\alpha=\beta$ and $\alpha \ne\beta$. In this regard, we observe that the
birefringent parameter $k$ involved in Eqs.(1.1) is real. However, the
symmetry algebra found for $\alpha=\beta$ does not depend on $k$. This fact
has suggested us to study the system also for imaginary values of $k$. In
such a situation, at least when $u=v$, $k$ can be interpreted as the loss
coefficient of the fiber ${\displaystyle{\cite{AG}}}$.
Anyway, we have obtained an interesting exact solution for $u \ne v$ as
well. This solution is derived from the Galilean boost. Another important
aspect of the symmetry reduction technique is the determination of the
infinitesimal operator which is responsible for the soliton profile and the
periodic configuration.

In Sec. 2 we outline the method of symmetry reduction and obtain the
symmetry algebra and the corresponding symmetry group related to Eqs.(1.1).
Sec. 3 contains examples of specific exact solutions, while in Sec. 4 some
concluding remarks are reported.

\section{The method of symmetry reduction}
\subsection{ a) The symmetry algebra}
The method of symmetry reduction (SR) consists of an application of the Lie
group theory to reduce Eqs. (1.1) to a system of ordinary differential
equations.

A fundamental step of the SR procedure is to obtain the Lie point
symmetries ${\displaystyle{\cite{OL}}}$ of Eqs. (1.1): in other words, the
symmetry algebra
$ {\cal L}$
 and the corresponding symmetry group ${\cal G}$ of the equations under
investigation.
Then, we can build up solutions that are invariants under some specific
subgroup ${\cal G}_0$ of ${\cal G}$. The SR can be carried out via the
determination of the invariants of  ${\cal G}_0$. Invariants are furnished
by the partial differential equations
$$V_{j} \; I(x,t,u,v,u^{*},v^{*}) = 0, \quad j=1,2,...,n,
\eqno(2.1)$$
where $\{V_{j}\}$ is a basis of the Lie algebra $ {\cal L}_{0}$ of
$ {\cal G}_{0}$ {\rm and} $n$ is the number of the independent elements (
infinitesimal operators) $V_{j}$ of  $ {\cal L}_{0}$. Once the invariants
related to  $ {\cal G}_{0}$ are known, Eqs.(1.1) can be written in terms of
them. In such a way, we are led to a set of reduced equations which may
 yield exact solutions to the original system (1.1).
The Lie point symmetries of Eqs. (1.1) can be found by resorting to the
standard technique outlined in ${\displaystyle{\cite{OL}}}$.
Precisely, let us introduce the vector
field
$$V = \xi_{1} \partial_{x} +  \xi_{2} \partial_{t} +  \xi_{3} \partial_{u}
+  \xi_{4} \partial_{u^{*}} +  \xi_{5} \partial_{v} +  \xi_{6}
\partial_{v^{*}},\eqno(2.2)$$
where $\xi_{j} \: (j= 1,2,...,6)$ are functions which depend in general on
$ x,t,u,u^{*},v,v^{*},$ and $\partial_{x} = \frac{\partial}{\partial x}$,
and so on.
A local group of transformations $G$ is a symmetry group for Eqs.(1.1) if
and only if
$$pr^{(2)} V[\Delta_{j}] =0, \quad \quad pr^{(2)} V[\Delta^{*}_{j}] =0, \quad
j= 1,2
, \eqno(2.3)$$
whenever $\Delta_{j}=0, \quad \Delta^{*}_{j}=0$  for every generator of $G$
, where $pr^{(2)}V$ is the second prolongation of $V$
${\displaystyle{\cite{OL}}}$.

The conditions (2.3) constitute a set of constraints in the form of partial
differential equations which enable us to obtain the coefficients $\xi_{j}$.
The calculations have been performed in part by using the symbolic language
REDUCE ${\displaystyle{\cite{SC}}}$. We have achieved the following results.

${\bf Case\: I : \alpha \ne \beta}$

The symmetry algebra is defined by four elements, namely:
$$ V_{1} = \partial_{x}, \quad  V_{2} = \partial_{t}, \quad
 V_{3} = i(u\partial_{u} -u^{*}\partial_{u^{*}}+ v\partial_{v} -
v^{*}\partial_{v^{*}}),$$
$$V_{4} = x\partial_{t} + \frac{i}{2}\,t\,(u\partial_{u} -u^{*}\partial_{u^{*}}
+ v\partial_{v} -v^{*}\partial_{v^{*}}), \eqno(2.4)$$
where the nonvanishing commutation relations are
$$[V_{1}, V_{4}] = V_{2}, \quad [V_{2}, V_{4}] = \frac{1}{2}V_{3}. \eqno(2.5)$$

${\bf Case\: II : \alpha = \beta}$

The symmetry algebra is of the {\em sl(3,C)} type. It is defined by eight
elements
: four of them coincide with the previous ones, while the others are given by
$$ V_{5} = x\partial_{x} + \frac{1}{2} t\partial_{t} -
 \frac{1}{2} (u\partial_{u} +u^{*}\partial_{u^{*}}
+ v\partial_{v} +v^{*}\partial_{v^{*}})+ ikx (v\partial_{u}
-v^{*}\partial_{u^{*}}
+ u\partial_{v} -u^{*}\partial_{v^{*}}), \eqno(2.6a)$$
$$ V_{6} = i(v\partial_{u} -v^{*}\partial_{u^{*}}
+ u\partial_{v} -u^{*}\partial_{v^{*}}), \eqno(2.6b)$$
$$ V_{7} = (
v\partial_{u} +v^{*}\partial_{u^{*}}
- u\partial_{v} -u^{*}\partial_{v^{*}})\cos{2kx}
- i(u\partial_{u} -u^{*}\partial_{u^{*}}
- v\partial_{v} +v^{*}\partial_{v^{*}})\sin{2kx}, \eqno(2.6c)$$
$$ V_{8} = (v\partial_{u} +v^{*}\partial_{u^{*}}
-u \partial_{v} -u^{*}\partial_{v^{*}})\sin{2kx}
+ i(u\partial_{u} -u^{*}\partial_{u^{*}}
- v\partial_{v} +v^{*}\partial_{v^{*}})\cos{2kx}. \eqno(2.6d)$$
The nonvanishing commutation relations fulfilled by $ V_{1},\dots, V_{8}$, are
(2.5) together with
$$[V_{1}, V_{5}] =  V_{1}+ k V_{6},\quad [ V_{1}, V_{7}] =-2k V_{8},\quad
[ V_{1}, V_{8}]=2k V_{7},$$
$$[V_{2},V_{5}]=\frac{1}{2}V_{2}, \qquad [V_{5},V_{4}]=\frac{1}{2}V_{4},
\eqno(2.7)$$
and
$$[ V_{6}, V_{7}]=2 V_{8},\quad [ V_{7}, V_{8}]=2 V_{6},\quad
[ V_{8}, V_{6}]=2 V_{7} \eqno(2.8)$$
The vector fields $V_{1},...,V_{8}$ are the generators of the infinitesimal
symmetries transformations of Eqs.(1.1).These are the coordinate
translations and the Galilean boost ($V_{4}$), which are common to both the
cases I and II. Moreover, for $\alpha =\beta$ Eqs. (1.1) admit the
additional symmetry {\em su(2,C)} which is expressed by the generators $V_{6},
\;V_{7},\:V_{8}$ satisfying the commutation rules (2.8)\footnote{ It is
noteworthy that for $\alpha \ne \beta$, the symmetry algebra turns out to
be independent from the parameter $k$. Conversely, for $\alpha=\beta$ and
$k$ such that ${\rm Im}k\ne0$, the symmetry algebra is defined by the
vector fields $V_{1},\,V_{2},\,V_{3},\,V_{4}\;{\rm and}\;V_{6}$.
In other words, the presence in
Eqs.(1.1) of a non-real coefficient in front of $u$ and $v$ changes the
symmetry algebra, which is no longer of the {\it sl(3,C)} type. This fact
is connected with the loss of the integrability property of Eqs.(1.1) for
$\alpha =\beta$ and $k$ such that ${\rm Im}k\ne0$.}.

\subsection{b) The group transformations}
By integrating the infinitesimal operators  $V_{1},...,V_{8}$, we provide
the group transformations that leave Eqs.(1.1) invariant.
These are, respectively:
 $$V_{1}: \quad \tilde{t} =t,\qquad\tilde{x} = x + \lambda,
\quad \tilde{u} =u,\quad
 \tilde{v} =v,\eqno(2.9a)$$
 $$V_{2}: \qquad \tilde{t} = t + \lambda, \quad \tilde{x} =x,\qquad
\tilde{u} =u,\quad \tilde{v} =v  ,\eqno(2.9b)$$
 $$V_{3}: \qquad \tilde{x} = x , \quad \tilde{t} =t,\quad
 \tilde{u} =u\: e^{i\lambda},\quad  \tilde{v} =v\:e^{i\lambda},\eqno(2.9c)$$

$$\begin{array}{ll}
V_{4}:\qquad
& \tilde{x}=x, \;\; \tilde{t}=t + \lambda \tilde{x}, \;\;
  \tilde{u}=u e^{\frac{i}{2}(\lambda \tilde{t} - \frac {1}{2} \lambda^{2}
  \tilde{x})},\\
& \tilde{v}=v e^{\frac{i}{2}(\lambda \tilde{t} - \frac {1}{2} \lambda^{2}
  \tilde{x})}, \end{array} \eqno(2.9d)$$

$$\begin{array}{ll}
V_{5}:\qquad
& \tilde{x}= e^{\lambda} x, \;\; \tilde{t}=e^{\lambda/2} t,\\

& \left( \begin{array}{c}
\tilde{u} \\  i\tilde{v}
 \end{array} \right)
=
  \sqrt{\frac{x }{\tilde{x}}}
\left(\begin{array}{cc} \cos{k(\tilde{x}-x)} & \sin{k(\tilde{x}-x)}\\
-\sin{k(\tilde{x}-x)} & \cos{k(\tilde{x}-x)}
\end{array}\right)
\left( \begin{array}{c}
u\\iv
\end{array} \right), \end{array} \eqno(2.9e)$$

\vspace{.5cm}

$$\begin{array}{ll}
V_{6}:\qquad
& \tilde{x}=  x, \;\; \tilde{t}=t,\\
& \left( \begin{array}{c}
\tilde{u}\\ \tilde{v} \end{array}\right)
= e^{i \lambda \sigma_{1}}
\left( \begin{array}{c}
u \\ v
\end{array} \right) ,

\end{array} \eqno(2.9f)$$

\vspace{.5cm}

$$\begin{array}{ll}
V_{7}:\qquad
& \tilde{x}=  x, \;\; \tilde{t}=t,\\

& \left( \begin{array}{c}
\tilde{u}\\ \tilde{v} \end{array}\right)
=e^{i\lambda(\cos{2kx}\; \sigma_{2} - \sin{2kx}\; \sigma_{3})}

\left( \begin{array}{c}
u \\ v
\end{array} \right),
\end{array} \eqno(2.9g)$$
\vspace{.5cm}

$$\begin{array}{ll}
V_{8}:\qquad
& \tilde{x}=  x, \;\; \tilde{t}=t,\\
&  \left( \begin{array}{c}
\tilde{u}\\ \tilde{v} \end{array}\right)
=e^{i\lambda( \sin{2kx}\; \sigma_{2} + \cos{2kx}\; \sigma_{3} )}

\left( \begin{array}{c}
u \\ v
\end{array} \right)
\end{array}, \eqno(2.9h)$$
\vspace{.5cm}

where $\lambda$ and $\sigma_{1}$, $\sigma_{2}$, $\sigma_{3}$ are the
group parameter and the Pauli matrices, respectively. In deriving (2.9), we
have used the initial conditions $\tilde{x}(\lambda)\mid_{\lambda=0}\,=x,\;\;
\;\tilde{t}(\lambda)\mid_{\lambda=0}\,=t,\;\;\;\tilde{u}(\lambda)
\mid_{\lambda=0}\,=u,\;\;\;\tilde{v}(\lambda)\mid_{\lambda=0}\,=v$.

Equations (2.9a)-(2.9h) tell us that if $u=f(x,t),\quad v=g(x,t)$ is a
solution of the system (1.1), so are
$$ u^{(1)}=f(x-\lambda,t), \;\;\; v^{(1)}=g(x-\lambda,t), \eqno(2.10a)$$
$$ u^{(2)}=f(x,t-\lambda), \;\;\; v^{(2)}=g(x,t-\lambda), \eqno(2.10b)$$
$$ u^{(3)}=e^{i\lambda}f(x,t), \;\;\; v^{(3)}=e^{i\lambda}g(x,t),\eqno(2.10c)$$

\vspace{.5cm}

$$ u^{(4)}=f(x,t-\lambda x)e^{\frac{i}{2}(\lambda t-\frac{1}{2}\lambda^2x)},$$
$$ v^{(4)}=g(x,t-\lambda x)e^{\frac{i}{2}(\lambda t-\frac{1}{2}\lambda^2x)},
\eqno(2.10d)$$

$$ u^{(5)}=e^{-\frac{\lambda}{2}}\{f(e^{-\lambda}x,e^{-\frac{\lambda}{2}}t)
\cos{[kx(e^{\lambda} -1)]}+ ig(e^{-\lambda}x,e^{-\frac{\lambda}{2}}t)
\sin{[kx(e^{\lambda} -1)]}\}, $$
$$ v^{(5)}=e^{-\frac{\lambda}{2}}\{g(e^{-\lambda}x,e^{-\frac{\lambda}{2}}t)
\cos{[kx(e^{\lambda} -1)]}+ if(e^{-\lambda}x,e^{-\frac{\lambda}{2}}t)
\sin{[kx(e^{\lambda} -1)]}\},$$
$$ \eqno(2.10e)$$

$$ u^{(6)}=f(x,t)\cos{\lambda} + ig(x,t)\sin{\lambda}, $$
$$ v^{(6)}= if(x,t)\sin{\lambda} + g(x,t)\cos{\lambda} , \eqno(2.10f)$$

$$ u^{(7)}= (\cos{\lambda} -i \sin{\lambda}\: \sin{2kx}) f(x,t) +
(\sin{\lambda}\: \cos{2kx}) g(x,t), $$
$$ v^{(7)}= (-\sin{\lambda}\: \cos{2kx}) f(x,t) + (\cos{\lambda} +i
\sin{\lambda}\: \sin{2kx})
g(x,t),  \eqno(2.10g)$$

$$ u^{(8)}=  (\cos{\lambda} +i \sin{\lambda}\: \cos{2kx}) f(x,t) +
(\sin{\lambda}\: \sin{2kx}) g(x,t) ,  $$
$$ v^{(8)}= (-\sin{\lambda} \:\sin{2kx}) f(x,t) + (\cos{\lambda} -i
\sin{\lambda}
\: \cos{2kx})
g(x,t).  \eqno(2.10h)$$

\section{Exact solutions}

As we have already mentioned, the method of symmetry reduction of a partial
differential equation amounts essentially to finding the invariants (
symmetry variables) of a given subgroup of the symmetry group admitted by
the equation under consideration. A basis set of invariants for the
generators $V_{j}$ can be obtained by solving Eq.(2.1). Alternatively,
one can resort to a direct equivalent procedure by using the group
transformations (2.9).

The invariants can be exploited to provide exact solutions to Eqs.(1.1).
 By way of example, in this Section we shall deal with the invariants
related to the symmetry operators   $a)\;\; V_{0}=\,V_{1} + V_{2}+ V_{3},\quad
b)\;\; V_{4},\quad c)\;\; V_{1}+ V_{4}\quad {\rm and}\;\; d)\;\; V_{5}$.

{\bf Case a)} A set of invariants related to $V_{0}$ is
$$y = \tilde{t}- \tilde{x} = t-x,$$
 $$U =  \tilde{u}e^{-i \tilde{x}}=ue^{-i x}, \quad
  W =  \tilde{v}e^{-i \tilde{x}}=ve^{-i x},$$
  $$U_{1} =  \tilde{u}e^{-i \tilde{t}}=ue^{-it},\quad
 W_{1} =  \tilde{v}e^{-i \tilde{t}}=ve^{-it}. \eqno(3.1)$$

Inserting, for instance,
$$u(x,t)=U(y)e^{ix}, \quad v(x,t)=W(y)e^{ix}, \eqno(3.2)$$
into Eqs.(1.1) (for $\alpha =\beta$) gives the pair of (ordinary) reduced
equations
$$iU^{\prime}+U-U^{\prime\prime}-kW-\alpha(|U|^{2}+|W|^{2})U=0, \eqno(3.3a)$$
$$iW^{\prime}+W-W^{\prime\prime}-kU-\alpha(|U|^{2}+|W|^{2})W=0, \eqno(3.3b)$$
where
$$U=U(y), \quad W=W(y), \quad U^{\prime}=\frac{dU}{dy}, \quad {\rm and} \;\;
W^{\prime}=\frac{dW}{dy}.$$
Now, let us look for solutions to Eqs.(3.3) of the type
$$U(y) = p(y)e^{i\gamma y}, \quad  W= q(y)e^{i\delta y} \eqno(3.4)$$
where $p,\;q$ are real functions of $y$ and $\gamma,\; \delta$ are real
constants. In doing so, Eqs.(3.3) yield
$$(2\gamma-1)p^{\prime}+kq \sin{(\delta-\gamma)y}=0,$$
$$(2\delta-1)q^{\prime}-kp \sin{(\delta-\gamma)y}=0, \eqno(3.5a)$$
$$p^{\prime\prime}+(\gamma-\gamma^{2}-1)p+kq \cos{(\delta-\gamma)y} +\alpha(
p^{2}+q^{2})p=0,$$
$$q^{\prime\prime}+(\delta-\delta^{2}-1)q+kp \cos{(\delta-\gamma)y} +\alpha(
p^{2}+q^{2})q=0. \eqno(3.5b)$$
$$(p^{\prime}=\frac{dp}{dy},\quad q^{\prime}=\frac{dq}{dy}).$$
Equations (3.5) produce some interesting configurations of Eqs.(1.1), such
as the soliton profile and solutions expressed in terms of the
elliptic Jacobi functions sn($\cdot$).

To this aim, let us choose $\gamma=\delta=\frac{1}{2}$. Then, by defining
$z=p+iq=\rho e^{i\pi/4}\;\;\;(\rho=|z|)$, we have
$$\rho^{\prime}{}^{2} = (\frac{3}{4}-k)\rho^{2}-\frac{\alpha}{2}\rho^{4}+c,
\eqno(3.6)$$
where
$\rho^{\prime}=\frac{d\rho}{dy}$ and $c$ is a constant of integration.
The soliton profile comes from (3.6) for $c$ = 0, by taking $\alpha>0,\;\;
k<\frac{3}{4}$. Precisely
$$p=q=\frac{1}{\sqrt{2}}\rho= \sqrt{(\frac{3}{4}-k)\frac{1}{\alpha}}\:
{\rm sech}\left[\sqrt{\frac{3}{4}-k}\:(y-y_{0})\right], \eqno(3.7)$$
where $y_{0}$ is an arbitrary constant. With the help of (3.7), from (3.2)
we obtain
$$ u=v=e^{\frac{i}{2}(t+x)}p(t-x),\eqno(3.8)$$
with $p(t-x)$ given by (3.7).

We notice that for $c=0$, and $\alpha<0,\;\;k>\frac{3}{4}$, Eq.(3.6) leads
to the solution
$$p=q=\frac{\rho}{\sqrt{2}}=\sqrt{(\frac{3}{4}-k)\frac{1}{\alpha}}\:
\sec{\left[\sqrt{k-\frac{3}{4}}\;(y-y_{0})\right]}.\eqno(3.9)$$
In this case, we loose the soliton character of the profile. (The onset of
the soliton depends on the parameters $\alpha$ and $k$).

Another interesting solution linked to the symmetry operator $V_{0}$
arises for $c=k-3/4-|\alpha|/2\:> 0$ and $\alpha<0,\;\;k>\frac{3}{4}$.
Indeed, from Eqs. (3.6), (3.4) and (3.2) we have
$$ u=v= \frac{1}{\sqrt{2}}e^{\frac{i}{2}(t-x)}{\rm sn}\left[ \sqrt{c}(t-x),
h\right], \eqno(3.10)$$
where sn$(\cdot)$ is the sinus elliptic function of modulus
$h= \sqrt{\frac{|\alpha|}{2c}}$.

{\bf Case b)} By using the basis of invariants
$$ \tilde{x}=x, \;\;\; U(x)= \tilde{u}e^{-i\tilde{t}^{2}/4\tilde{x}}
= ue^{-it^{2}/4x},$$
$$  W(x)= \tilde{v}e^{-i\tilde{t}^{2}/4\tilde{x}}
= ve^{-it^{2}/4x},\eqno(3.11)$$
associated with the vector field $V_{4}$, from Eqs.(1.1) we find the pair
of reduced equations
$$ i\left( U^{\prime}+\frac{1}{2x}U\right)+kW+\left(\alpha|U|^{2}+
\beta|W|^{2}\right)U=0, \eqno(3.12a)$$
$$ i\left( W^{\prime}+\frac{1}{2x}W\right)+kU+\left(\alpha|W|^{2}+
\beta|U|^{2}\right)W=0,  \eqno(3.12b)$$
where $U^{\prime}=\frac{dU}{dx},\;\;\;W^{\prime}=\frac{dW}{dx}$. For
$\alpha=\beta$ and assuming that $k$ is real, Eqs.(3.12) can be solved in the
following way. First, let us
divide (3.12a) and (3.12b) by U and W, respectively. Second, let us subtract
the
resulting equation corresponding to (3.12a) from that corresponding to
(3.12b).
Then, we obtain
$$ i\left( WU^{\prime} - UW^{\prime}\right) + k\left(W^{2}-U^{2}\right)=0.
\eqno(3.13)$$
By introducing now
$$ U= \frac{1}{2}(A+B),\quad  W= \frac{1}{2}(A-B) \eqno(3.14)$$
into Eq.(3.13), we can determine $B$ in terms of $A$, namely
$$ B=Ae^{-2i(kx+\delta_{0})}, \eqno(3.15)$$
where $\delta_{0}$ is a constant of integration. Hence, putting the
quantities (3.14) into Eq.(3.12a) and taking account of (3.15), after some
manipulations we arrive at the solutions
$$ u= \frac{a}{\sqrt{x}}e^{i(\frac{t^{2}}{4x}-\delta_{0})}\;x^{i\alpha a^{2}}
\; \cos{(kx+\delta_{0})}, \eqno(3.16a)$$
$$ v= i\frac{a}{\sqrt{x}}e^{i(\frac{t^{2}}{4x}-\delta_{0})}\;x^{i\alpha a^{2}}
\; \sin{(kx+\delta_{0})}, \eqno(3.16b)$$
where $a$ is a real constant. The physical
role of these configurations remains to be explored. Here we  limit
ourselves to observe that the 'mass density' $|u|^{2}+
|v|^{2}$ is of the Coulomb-type in the $x-$variable and is time independent.
Moreover, the quantities (3.16) resemble the singular solution to the
nonlinear Schr\"{o}dinger equation expressed by formula (3.10) of ref.9.

An interesting class of exact solutions related to the generator $V_{4}$
can be found for $\alpha \ne \beta$ and $k=ik_{0}$, where $k_{0}$ is a
real number.

In this case, let us look for solutions to Eqs.(3.12) of the type
$$U=\rho\,e^{i\theta},\quad W=\rho\,e^{i\gamma}, \eqno(3.17)$$
where $\rho,\;\theta\;{\rm and}\; \gamma$ are real functions of $x$.
Inserting (3.17) into Eqs.(3.12) yields
$$\frac{\rho^{\prime}}{\rho}\,+\,\frac{1}{2x}\,+\,k_{0}\,\cos{(\gamma-
\theta)}\,=0, \eqno(3.18a)$$
$$\theta^{\prime}\,+\,k_{0}\sin{(\gamma-\theta)}\,-\,(\alpha+\beta)\,\rho^{2}
=0\,, \eqno(3.18b).$$
$$\gamma^{\prime}\,-\,k_{0}\sin{(\gamma-\theta)}\,-\,(\alpha+\beta)\,\rho^{2}
=0\,, \eqno(3.18c)$$
By integrating this system and using (3.11), we give the following pair of
exact solutions to Eqs.(1.1):
$$u\,=\rho\,e^{i(\frac{t^{2}}{4x}\,+\,\theta)},\quad v\,=\, u\,e^{i
\arcsin{[{\rm sech}2k_{0}(x_{0}\,-\,x)]}}, \eqno(3.19)$$
where
$$\rho\,=\,\sqrt{\frac{\cosh{2k_{0}(x_{0}-x)}}{x}},$$
$$ \theta\,=\,\frac{1}{2}
\arcsin{[\cosh{2k_{0}(x_{0}-x)}]}\,+\,(\alpha\,+\,\beta)\,c^{2}\,
\int^{x}_{x_{1}}\frac{\cosh{2k_{0}(x_{0}-x^{\prime})}}{x^{\prime}}\;
dx^{\prime}\;, \eqno(3.20)$$
and $c,\,x_{0},\,x_{1},\,$ are real constants.

At this point, let us deal with the special choice $\theta\,=\, \gamma$.
Then, Eqs.(3.18) lead to the solution
$$u\, =\,v=\, c\,\frac{e^{-k_{0}x}}{\sqrt{x}}\,e^{i\frac{t^{2}}{4x}}\,
e^{i\;(\alpha +\beta)\,c^{2}
\int^{x}_{x_{1}}\frac{e^{-2k_{0}x^{\prime}}}{x^{\prime}}\,dx^{\prime}}\;,
\eqno(3.21)$$
where $c\; {\rm and}\; x_{1}$ are constants. (For $x_{1}\,\rightarrow
\infty,$ the
integral at the r.h.s. of (3.21) becomes $-{\rm E}_{1}(2kx)$, where
${\rm E}_{1}(\cdot)$
 is the exponential integral function ${\displaystyle{\cite{HA}}}$).
The 'mass density' corresponding to the solutions (3.20) and (3.21)
is
$$|u|^{2}\,+\,|v|^{2}\,=\, 2c^{2}\frac{\cosh2k_{0}(x_{0}-x)}{x}\quad {\rm and}
\quad |u|^{2}\,=\,|v|^{2}\,=\,c^{2}\frac{e^{-2k_{0}x}}{x}\,.\eqno(3.22)$$
We point out that for $k=0,$ the mass densities (3.22) behave as $1/x.$
Consequently, the presence of the parameter $k=ik_{0}$ induces a
change of the Coulomb-like mass density.

{\bf Case c)} A basis of invariants related to the symmetry operator $V_{1}+
V_{4}$ is given by
$$ \eta\, =\, \tilde{t}\,-\,\frac{1}{2}\,\tilde{x}^{2}\, =
t\,-\,\frac{1}{2}\,x^{2},$$
$$
U(\eta)\,=\,\tilde{u}\,e^{-\frac{i}{2}\tilde{x}(\eta+\frac{1}{6}\tilde{x}^{2})}\,=
\,u\,e^{-\frac{i}{2}x(\eta+\frac{1}{6}x^{2})},$$
$$
W(\eta)\,=\,\tilde{v}\,e^{-\frac{i}{2}\tilde{x}(\eta+\frac{1}{6}\tilde{x}^{2})}\,=
\,v\,e^{-\frac{i}{2}x(\eta+\frac{1}{6}x^{2})}. \eqno(3.23)$$
By assuming $\alpha=\beta,$ and setting (3.23) into Eqs.(1.1),
we get the reduced system
$$U^{\prime\prime}\,-\,\frac{1}{2}\,\eta\,U\,+\,k\,W\,+\,\alpha\,(|U|^{2}\,
+\,|W|^{2})\,U\,=\,0, \eqno(3.24a)$$
$$W^{\prime\prime}\,-\,\frac{1}{2}\,\eta\,W\,+\,k\,U\,+\,\alpha\,(|W|^{2}\,
+\,|U|^{2})\,W\,=\,0, \eqno(3.24b).$$
where $U^{\prime}\,=\,\frac{dU}{d\eta}.$
By requiring that $U\; {\rm and}\; W$ are real functions and $U\,=\,W$,
Eqs.(3.24) lead to a special case of the second Painlev\'e equation
${\displaystyle{\cite{IN}}}$, i.e.
$$\frac{d^{2}\psi}{dz^{2}}=z\psi\,+\,2\psi^{3}, \eqno(3.25)$$
where $z\,=\,2^{\frac{2}{3}}\,(\frac{1}{2}\eta -k)$ and
$ \psi\,=\,2^{\frac{1}{3}}\;\sqrt{-\alpha}\,U.$

{\bf Case d)} A set of invariants arising from the generator $V_{5}$ is
$$  \xi=\,\frac{\tilde{x}}{\tilde{t}^2}\,=\,\frac{x}{t^2},$$
$$I=\,\sqrt{\tilde{x}}(\tilde{u} \sin{k\tilde{x}} + i\tilde{v}
\cos{k\tilde{x}})\,=\,\sqrt{x}(u \sin{kx} + iv \cos{kx}),$$
$$ J=\,\sqrt{\tilde{x}}(\tilde{u} \cos{k\tilde{x}} - i\tilde{v}
\sin{k\tilde{x}})\,=\,\sqrt{x}(u \cos{kx} - iv \sin{kx}), \eqno(3.26)$$
from which
$$ K=\,\tilde{x}(|\tilde{u}|^{2}  + |\tilde{v}|^{2} )\,=
\,x(|u|^{2}  + |v|^{2} ). $$
Equations (3.26) imply
$$u=\,\frac{1}{\sqrt{x}}\left(I\sin kx +J\cos kx  \right), \eqno(3.27a)$$
$$v=\,\frac{i}{\sqrt{x}}\left(-I\cos kx +J\sin kx  \right). \eqno(3.27b)$$
Then, substitution from (3.27) into Eqs.(1.1) (for $\alpha=\beta$) yields
$$\frac{1}{2i}\left(I-2\xi I^{\prime}\right)\,+\,6\xi^{2}I^{\prime}\,+\,
4\xi^{3}I^{\prime\prime}\,+\,\alpha\left(|I|^{2}+|J|^{2}\right)I=0,
\eqno(3.28a)$$
$$\frac{1}{2i}\left(J-2\xi J^{\prime}\right)\,+\,6\xi^{2}J^{\prime}\,+\,
4\xi^{3}J^{\prime\prime}\,+\,\alpha\left(|I|^{2}+|J|^{2}\right)J=0,
\eqno(3.28b)$$
where $I=I(\xi),\;J=J(\xi),\;I^{\prime}=\frac{dI}{d\xi}\quad{\rm and}\;
J^{\prime}=\frac{dJ}{d\xi}.$
A simple solution to the system (3.28) can be found supposing that $I(\xi)$
and $J(\xi)$ are real functions. In such a case, Eqs.(3.28) can be easily
integrated. We have
$$u\,=\,\sqrt{\frac{-2}{\alpha}}\;\frac{\sin(kx+\phi)}{t},$$
$$v\,=\,-i\sqrt{\frac{-2}{\alpha}}\;\frac{\cos(kx+\phi)}{t},\eqno(3.29)$$
from (3.27), where $\phi$ is a constant and $\alpha$ may take both positive
and negative values.

\section{ Conclusions}
We have found some exact solutions to the system of equations (1.1)
describing the propagation of waves in birefringent optical fibers. These
equations have been analyzed in the framework of the Lie group theory. We
have determined the associated symmetry algebra and the corresponding group
transformations that leave Eqs.(1.1) invariant. Explicit configurations have
been obtained both in the integrable and in the nonintegrable case. The
subgroup (of the symmetry group) responsible for the soliton profile has
been provided. Not all the configurations arising from the symmetry
approach have an evident physical meaning. This problem remains open.
Notwithstanding, the knowledge of exact solutions to Eqs.(1.1) may be used
with benefit as a guide for the development of perturbative techniques or
for numerical calculations. Furthermore, the existence of exact solutions
could be a challenge for trying to create new experimental patterns.

\end{document}